\documentclass[aps,prl,twocolumn,superscriptaddress,showpacs]{revtex4}

\usepackage[dvips]{graphicx}
\usepackage{xspace}

\graphicspath{{./figures/}}

\begin{document}

% definitions here
\newcommand{\superk}    {Super-Kamiokande\xspace}       

\newcommand{\nue}       {$\nu_{e}$\xspace}
\newcommand{\numu}      {$\nu_{\mu}$\xspace}
\newcommand{\nutau}     {$\nu_{\tau}$\xspace}
\newcommand{\nusterile} {$\nu_{sterile}$\xspace}
\newcommand{\mutau}     {$\nu_\mu \rightarrow \nu_{\tau}$\xspace}
\newcommand{\musterile} {$\nu_\mu \rightarrow \nu_{sterile}$\xspace}
\newcommand{\dms}       {$\Delta m^2$\xspace}
\newcommand{\sstt}      {$\sin^2 2 \theta$\xspace}

\newcommand{\Rnqe}      {$R_{nqe}$\xspace}

\title{Indications of Neutrino Oscillation in
  a 250 km Long-baseline Experiment}

% Affiliation definitions and then the authors

\newcommand{\SNU}{\affiliation{ Department of Physics, Seoul National
    University, Seoul 151-742, KOREA}}

\newcommand{\Kobe}{\affiliation{ Kobe University, Kobe, Hyogo
    657-8501, JAPAN}}

\newcommand{\UW}{\affiliation{ Department of Physics, University of
    Washington, Seattle, WA 98195-1560, USA }}

\newcommand{\UCI}{\affiliation{ Department of Physics and Astronomy,
    University of California, Irvine, Irvine, CA 92697-4575, USA }}

\newcommand{\CNU}{\affiliation{ Department of Physics, Chonnam
    National University, Kwangju 500-757, KOREA}}

\newcommand{\ICRR}{\affiliation{ Institute for Cosmic Ray Research,
    University of Tokyo, Kashiwa, Chiba 277-8582, JAPAN}}

\newcommand{\Kyoto}{\affiliation{ Department of Physics, Kyoto
    University, Kyoto 606-8502, JAPAN}}

\newcommand{\Tohoku}{\affiliation{ Research Center for Neutrino
    Science, Tohoku University, Sendai, Miyagi 980-8578, JAPAN}}

\newcommand{\KEK}{\affiliation{ Institute of Particle and Nuclear
    Studies, KEK, Tsukuba, Ibaraki 305-0801, JAPAN }}

\newcommand{\SUNY}{\affiliation{ Department of Physics and Astronomy,
    State University of New York, Stony Brook, NY 11794-3800, USA}}

\newcommand{\Okayama}{\affiliation{ Department of Physics, Okayama
    University, Okayama, Okayama 700-8530, JAPAN}}

\newcommand{\BU}{\affiliation{ Department of Physics, Boston
    University, Boston, MA 02215, USA}}

\newcommand{\Hawaii}{\affiliation{ Department of Physics and
    Astronomy, University of Hawaii, Honolulu, HI 96822, USA}}

\newcommand{\Warsaw}{\affiliation{ Institute of Experimental Physics,
    Warsaw University, 00-681 Warsaw, POLAND }}

\newcommand{\SINS}{\affiliation{ A. Soltan Institute for Nuclear
    Studies, 00-681 Warsaw, POLAND}}

\newcommand{\KU}{\affiliation{ Department of Physics, Korea
    University, Seoul 136-701, KOREA}}

\newcommand{\Niigata}{\affiliation{ Department of Physics, Niigata
    University, Niigata, Niigata 950-2181, JAPAN}}

\newcommand{\Dongshin}{\affiliation{ Department of Physics, Dongshin
    University, Naju 520-714, KOREA}}

\newcommand{\MIT}{\affiliation{ Department of Physics, Massachusetts
    Institute of Technology, Cambridge, MA 02139, USA}}

\newcommand{\TSU}{\affiliation{ Department of Physics, Tokyo
    University of Science, Noda, Chiba 278-0022, JAPAN}}

% The alternative affilations will show up in the references

\newcommand{\now}{\altaffiliation{For current affiliations see
    http://neutino.kek.jp/present-addresses0211.ps}}

% not used for now

\newcommand{\Osaka}{\altaffiliation{ Department of Physics, Osaka
    University, Toyonaka, Osaka 560-0043, JAPAN}}

\newcommand{\Tokai}{\altaffiliation{ Department of Physics, Tokai
    University, Hiratsuka, Kanagawa 259-1292, JAPAN}}

\newcommand{\UPnow}{\altaffiliation{ Present address: University of
  Pittsburgh, Pittsburgh, PA 15260, USA}}

\newcommand{\Miyakyonow}{\altaffiliation{ Present address: Department of
    Physics, Miyagi University of Education, Sendai 980-0845, JAPAN}}

\newcommand{\Hillnow}{\altaffiliation{ Present address: California State
    University, Dominghez Hills, USA}}

\newcommand{\Jangnow}{\altaffiliation{ Present address: Seokang College,
    Kwangju, 500-742, KOREA}}

\newcommand{\Kainow}{\altaffiliation{ Present address: Department of
    Physics, University of Utah, Salt Lake City, UT 84112, USA}}

\newcommand{\Marunow}{\altaffiliation{ Present address: The Enrico Fermi
    Institute, University of Chicago, Chicago, IL 60637, USA}}

\newcommand{\Maugernow}{\altaffiliation{ Present address: California
    Institute of Technology, California 91125, USA}}

\newcommand{\Nagoyanow}{\altaffiliation{ Present address: Department of
    Physcs, Nagoya University, Nagoya, Aichi 464-8602, JAPAN}}

\newcommand{\DankaSup}{\altaffiliation{ Supported by the Polish Committee
    for Scientific Research}}

\author{M.H.Ahn}\SNU
\author{S.Aoki}\Kobe
\author{H.Bhang}\SNU
%\author{S.Boyd}\UW
\author{S.Boyd}\now\UW
\author{D.Casper}\UCI
\author{J.H.Choi}\CNU
\author{S.Fukuda}\ICRR
%\author{Y.Fukuda}\ICRR
\author{Y.Fukuda}\now\ICRR
\author{W.Gajewski}\UCI
\author{T.Hara}\Kobe
\author{M.Hasegawa}\Kyoto
\author{T.Hasegawa}\Tohoku
\author{Y.Hayato}\KEK
%\author{J.Hill}\SUNY
\author{J.Hill}\now\SUNY
\author{A.K.Ichikawa}\KEK
\author{A.Ikeda}\Okayama
\author{T.Inagaki}\Kyoto
\author{T.Ishida}\KEK
\author{T.Ishii}\KEK
\author{M.Ishitsuka}\ICRR
\author{Y.Itow}\ICRR
\author{T.Iwashita}\Kobe
\author{H.I.Jang}\now\CNU
%\author{H.I.Jang}\CNU
\author{J.S.Jang}\CNU
\author{E.J.Jeon}\KEK
\author{C.K.Jung}\SUNY
\author{T.Kajita}\ICRR
\author{J.Kameda}\ICRR
\author{K.Kaneyuki}\ICRR
\author{I.Kato}\Kyoto
\author{E.Kearns}\BU
\author{A.Kibayashi}\Hawaii
\author{D.Kielczewska}\Warsaw\SINS
\author{K.Kobayashi}\ICRR
\author{B.J.Kim}\SNU
\author{C.O.Kim}\KU
\author{J.Y.Kim}\CNU
\author{S.B.Kim}\SNU
\author{T.Kobayashi}\KEK
\author{M.Kohama}\Kobe
\author{Y.Koshio}\ICRR
\author{W.R.Kropp}\UCI
\author{J.G.Learned}\Hawaii
\author{S.H.Lim}\CNU
\author{I.T.Lim}\CNU
\author{H.Maesaka}\Kyoto
\author{K.Martens}\now\SUNY
%\author{K.Martens}\SUNY
\author{T.Maruyama}\now\KEK
%\author{T.Maruyama}\KEK
\author{S.Matsuno}\Hawaii
\author{C.Mauger}\now\SUNY
%\author{C.Mauger}\SUNY
\author{C.Mcgrew}\SUNY
\author{S.Mine}\UCI
\author{M.Miura}\ICRR
\author{K.Miyano}\Niigata
\author{S.Moriyama}\ICRR
\author{M.Nakahata}\ICRR
\author{K.Nakamura}\KEK
\author{I.Nakano}\Okayama
\author{F.Nakata}\Kobe
\author{T.Nakaya}\Kyoto
\author{S.Nakayama}\ICRR
\author{T.Namba}\ICRR
\author{K.Nishikawa}\Kyoto
\author{S.Nishiyama}\Kobe
\author{S.Noda}\Kobe
\author{A.Obayashi}\ICRR
\author{A.Okada}\ICRR
\author{T.Ooyabu}\ICRR
\author{Y.Oyama}\KEK
\author{M.Y. Pac}\Dongshin
\author{H.Park}\SNU
\author{M.Sakuda}\KEK
\author{N.Sakurai}\ICRR
\author{N.Sasao}\Kyoto
\author{K.Scholberg}\MIT
\author{E.Sharkey}\SUNY
\author{M.Shiozawa}\ICRR
\author{H.So}\SNU
\author{H.W.Sobel}\UCI
\author{A.Stachyra}\UW
\author{J.L.Stone}\BU
\author{Y.Suga}\Kobe
\author{L.R.Sulak}\BU
\author{A.Suzuki}\Kobe
\author{Y.Suzuki}\ICRR
\author{Y.Takeuchi}\ICRR
\author{N.Tamura}\Niigata
\author{T.Toshito}\now\ICRR
%\author{T.Toshito}\ICRR
\author{Y.Totsuka}\ICRR\KEK
\author{M.R.Vagins}\UCI
\author{C.W.Walter}\BU
\author{R.J.Wilkes}\UW
\author{S.Yamada}\ICRR
\author{S.Yamamoto}\Kyoto
\author{C.Yanagisawa}\SUNY
\author{H.Yokoyama}\TSU
\author{J.Yoo}\SNU
\author{M.Yoshida}\KEK
\author{J.Zalipska}\SINS

\collaboration{The K2K Collaboration}\noaffiliation

%%% Local Variables: 
%%% mode: latex
%%% TeX-master: "oscillation"
%%% End: 

\date{\today}

\begin{abstract}
  The K2K experiment observes indications of neutrino oscillation: 
  a reduction of
  $\nu_\mu$ flux together with a distortion of the energy
  spectrum.  Fifty-six beam neutrino events are observed in
  Super-Kamiokande~(SK), 250~km from the neutrino production point,
  with an expectation of $80.1^{+6.2}_{-5.4}$.  Twenty-nine one
  ring $\mu$-like events are used to reconstruct the neutrino energy
  spectrum, which is better matched to the expected spectrum with
  neutrino oscillation than without.  The probability that the
  observed flux at SK is explained by statistical fluctuation
  without neutrino oscillation is less than 1\%. 
\end{abstract}


\maketitle

%\section{Introduction}
%\label{sec:introduction}

Recent atmospheric and solar neutrino data indicate the
existence of neutrino oscillation and therefore the existence of
neutrino mass~\cite{Fukuda:1998mi,Fukuda:2002ds,Ahmad:2002de}.  The
zenith angle distribution of atmospheric neutrinos observed by \superk
(SK) shows a clear deficit of upward-going muon neutrinos, which is
well explained by two-flavor \numu-\nutau oscillations with 
\dms around $\rm 3 \times 10^{-3} eV^2$, and ${\rm sin^22\theta}$
close to or equal to unity.

The KEK to Kamioka long-baseline neutrino oscillation experiment
(K2K)~\cite{Ahn:2001cq} uses an accelerator-produced neutrino beam
with a neutrino flight distance of 250~km to probe the same \dms
region as that explored with atmospheric neutrinos.  The neutrino beam
is produced by a 12~GeV proton beam from the KEK proton
synchrotron.  After the protons hitting an aluminum target,
the produced positively charged particles, mainly pions, are focused
by a pair of pulsed magnetic horns~\cite{Ieiri:1997td}.
The neutrinos produced from the decays of these
particles are 98\% pure muon neutrinos with a mean energy
of 1.3~GeV.  

This analysis is based on data taken from June 1999 to July 2001,
corresponding to ${\rm 4.8\times 10^{19}}$ protons on target (POT).
The pion momentum and angular distributions downstream of the second horn are
occasionally measured with a gas-Cherenkov detector~(PIMON)~\cite{Maruyama}
in order to verify the beam Monte Carlo~(MC) simulation and to estimate 
the errors on the flux prediction at SK.  
The direction of the beam is monitored on a spill-by-spill basis by 
observing the profile of the muons from 
the pion decays with a set of ionization chambers and silicon
pad detectors located just after the beam dump.  The neutrino beam itself
is measured in a set of near neutrino detectors~(ND) located
300~m from the proton target.  The measurements made at the ND are
used to verify the stability and the direction of the beam, and to
determine the flux normalization and the energy spectrum before the
neutrinos travel the 250~km to SK.  
The flux at SK is estimated from the flux of the ND by multiplying
the Far/Near~(F/N) ratio, the ratio of fluxes between the far
detector (SK) and ND, to that of the ND.

Since both a suppression in the number of events and a distortion of
the spectrum are expected for  neutrinos which travel a fixed
path length in the presence of oscillations, both the number of
observed events and the spectral shape information at SK are compared
with expectation.  All of the beam-induced neutrino events observed
within the fiducial volume of SK are used to measure the overall
suppression of flux.  In order to study the spectral distortion,
1 ring $\mu$-like events~($1R \mu$) are selected to enhance the
fraction of charged-current (CC) quasi-elastic
~(QE)~interactions ($\nu_\mu + n \rightarrow \mu + p$).  Only the muon is
visible in these reactions since the proton momentum is typically
below Cherenkov threshold.  The energy of the parent neutrino can be
calculated by using the observed momentum of the muon, assuming 
QE interactions, and neglecting Fermi momentum:
\begin {eqnarray}
E_\nu ^{\rm rec}=\frac {m_NE_\mu-m^2_\mu/2} {m_N-E_\mu+P_\mu\cos\theta_\mu}, 
\label{eq:Enurec}
\end {eqnarray}

\noindent
where $m_N$, $E_\mu$, $m_\mu$, $P_\mu$ and $\theta_\mu$ are 
the nucleon mass, muon energy, the muon mass, the muon momentum and the 
scattering angle relative to the neutrino beam direction, respectively.  

%%% Local Variables: 
%%% mode: latex
%%% TeX-master: "oscillation"
%%% End: 

%\section{Measurement of Neutrino Flux and Spectrum at the near site}
%\label{sec:flux-measurement}

The ND is comprised of two detector systems: a 1~kiloton water
Cherenkov detector (1KT) and a fine-grained detector (FGD) system.
The flux normalization is measured by the 1KT to estimate the expected
number of events at SK.  Since the 1KT has the same detector
technology as SK, most of systematic uncertainties on the measurement
are canceled.  The energy spectrum is measured by analyzing the muon
momentum and angular distributions in both detector systems.  The
1KT has a high efficiency for reconstructing the momentum of muons
below 1~${\rm GeV}/c$, and full 4$\pi$ coverage in solid angle.
However, the 1KT has little efficiency for reconstructing muons with
momentum above 1.5~${\rm GeV}/c$ since they exit the detector.
The FGD, on the other hand, has high efficiency for measuring muons
above 1~${\rm GeV}/c$, and the two complementary detectors are thus
able to completely cover the relevant energy range.
 
%\subsection{1 kiloton}
In the 1KT analysis, a cylindrical fiducial volume of 25~tons 
oriented along the beam direction is used.  
Event selection criteria for the flux
normalization are the same as those in reference~\cite{Ahn:2001cq}.
Events which deposit more than $\sim 100$~MeV of energy are used for
the measurement of the integrated flux.  The measurement has a 5\%
systematic uncertainty, of which the largest contribution comes from
the vertex reconstruction~\cite{Ahn:2001cq}.
For the spectrum measurement, further cuts are imposed in order
to select $1R\mu$ events which stop in the detector. Among the events
selected for the flux normalization measurement, 53\% of the events
have 1 ring.  The events with a muon exiting from the detector are
effectively eliminated by requiring the maximum charge of any PMT to
be less than 200~photo-electrons; 68\% of the 1 ring events remain
after this requirement.  The largest systematic uncertainty for the
spectrum measurement is an uncertainty on the energy scale.
The energy scale is understood to within
$^{+2}_{-3}$~\%, which is confirmed with both cosmic-ray muons and
beam-induced $\pi^0$s.

%\subsection{Fine-Grained Detector}

The FGD is comprised of a scintillating fiber and water
detector~(SciFi)~\cite{Suzuki:2000nj}, 
a lead-glass calorimeter~(LG), and a muon range
detector~(MRD)~\cite{Ishii:2002nj}.  In the FGD analysis, events 
containing one or two tracks with vertex within the 5.9~ton fiducial 
volume of the SciFi are used.
The track finding efficiency is 70\% for
a track passing through three layers of scintillating fiber and close
to 100\% for more than 5 layers~\cite{Kim:2002pa}. 
Three layers is the minimum track length required in this analysis.
Events which have at
least one track passing into the MRD are chosen in order to select
$\nu_\mu$-induced CC interactions.  The momentum of the track is
measured by its range through the SciFi, LG, and MRD, with accuracy of
3\%.

\begin{figure}[htbp!]
  \centering  
  
  \centerline{ \includegraphics[width=4.25cm]{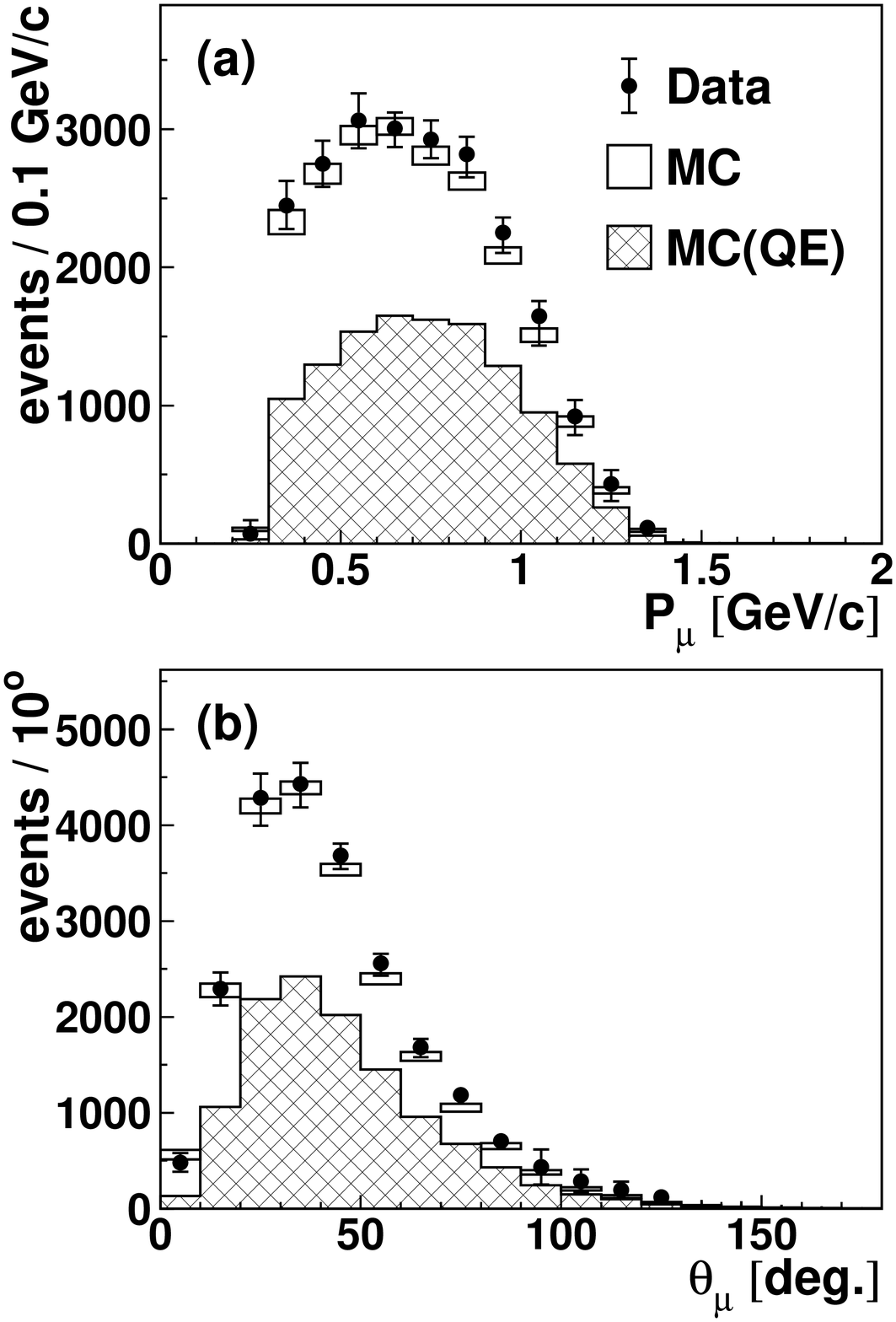}
               \includegraphics[width=4.25cm]{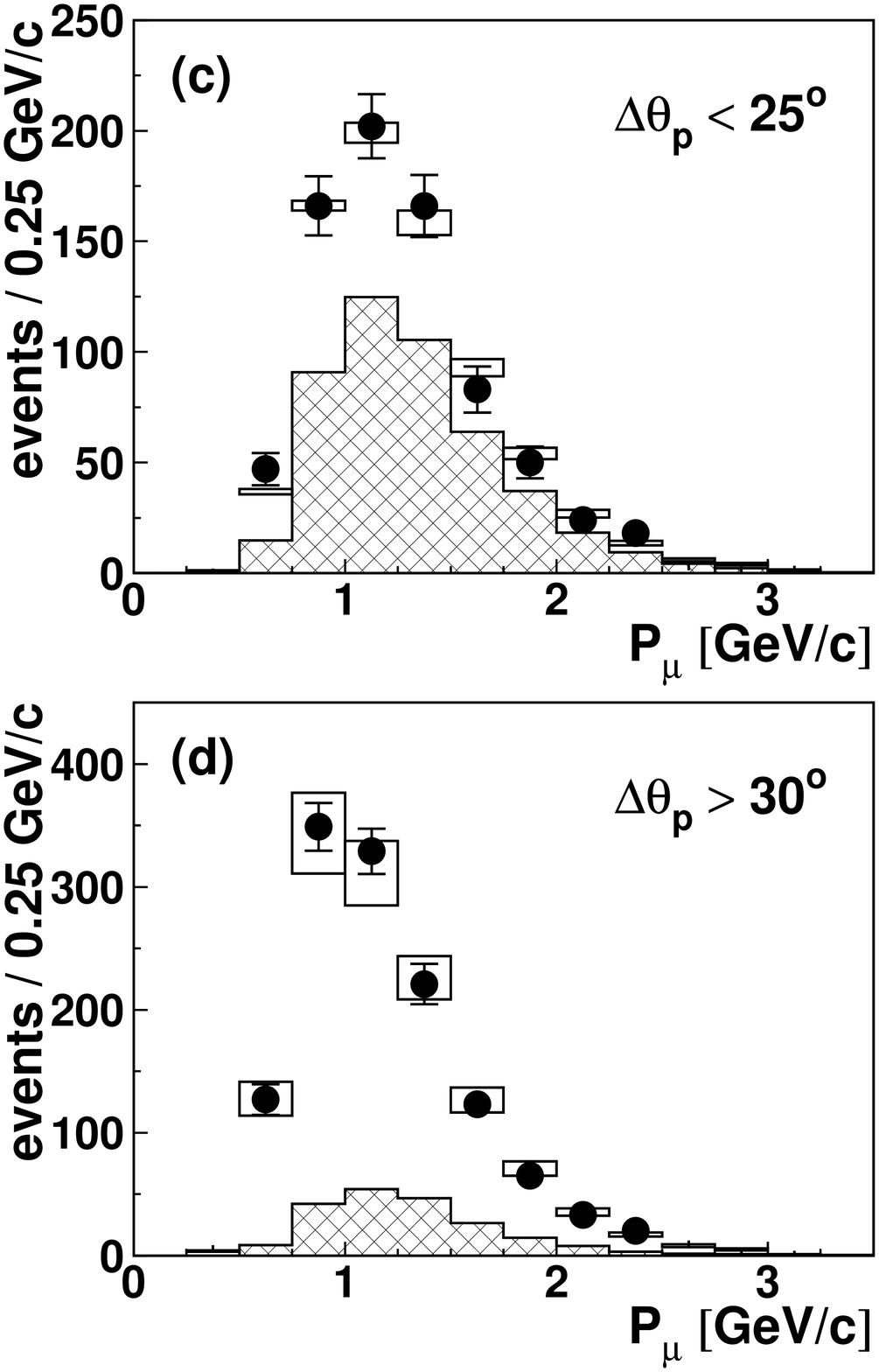}}
  \caption{(a) The muon momentum distribution of the 1KT $1R \mu$ sample,
           (b) the angular distribution of the 1KT $1R \mu$ sample,
           (c) the muon momentum distribution of the SciFi QE
               enhanced sample, and 
           (d) that of the SciFi non-QE enhanced sample.
           The crosses are data and the boxes are MC simulation with the
           best fit parameters. The hatched histogram shows the QE 
           events estimated by MC simulation.}
  \label{fig:kt}
\end{figure}

If the proton produced in the QE interaction has a momentum greater
than 600~MeV/$c$, its track may also be reconstructed.  In the
case where a second track is visible, the kinematic information is
used to enhance the fraction of QE events in the sample.  Assuming
QE interaction, the direction of the proton can be
predicted from the muon momentum. The QE enhanced sample is selected by
requiring that the direction of the second track agrees with the
prediction to within 25 degrees.  Events where the direction of the
second track differs from the prediction by more than 30 degrees are
put into a non-QE enhanced sample.  In the QE enhanced sample, 62\%
of the events are estimated to be QE events from the MC simulation.  In the
non-QE enhanced sample, $82$\% of events are estimated to come from
interactions other than QE.
The SciFi events are divided into three event categories: 1-track,
2-track QE enhanced, and 2-track non-QE enhanced samples.  

%\subsection{Merged Fit}
The 2-dimensional distributions of the muon momentum versus angle with
respect to the beam direction of four event categories (the 1KT event
sample and the three SciFi event samples) are used to measure the
neutrino spectrum.  A $\chi^2$-fitting method is used to compare these
data against the MC expectation.  The neutrino spectrum is divided
into 8 energy bins as defined in Table~\ref{tbl:SKspecerr}.
During the fit, the flux in each energy bin is re-weighted relative to
the values in the beam MC. These weights are normalized relative
to 1.00 for the $E_\nu=1.0$$-$1.5~GeV bin. 
An overall normalization is included as a free parameter in the fit.
The parameter \Rnqe is used to re-weight the ratio between the
QE and non-QE cross-section relative to the MC simulation.
The systematic uncertainties of the ND are 
incorporated into the fitting parameters.  They are the energy
scales, the track finding efficiencies, and the detector thresholds.
In addition, the spectrum measurement by PIMON is used as a constraint
on the re-weighting factors.

The value of $\chi^2$ is 227.2/197 d.o.f. at the best fit point. All
the parameters including the detector systematics are found to lie
within their expected errors.
The best fit values of the flux re-weighting factors are shown in
Table~\ref{tbl:SKspecerr}.
The muon momentum and angular distributions of $1R \mu$ events in the 1KT, and
the muon momentum distributions of the 2-track QE enhanced and
non-QE enhanced events in SciFi are overlaid with the re-weighted MC
in Figure~\ref{fig:kt}.  The fit result agrees well with the data.
The errors of the measurement are provided in the form of an error
matrix.  Correlations between the parameters of the fit are taken
into account in the oscillation analysis using this matrix.  The
diagonal elements in the error matrix are shown in Table~\ref{tbl:SKspecerr}.

\begin{table}[htbp!]
\caption{The central values of the flux re-weighting parameters
  for the spectrum fit at ND ($\Phi_{\rm{ND}}$) and
  the percentage size of the energy dependent systematic errors
  on $\Phi_{\rm{ND}}$, F/N ratio, and $\epsilon_{\rm{SK}}$.
  The re-weighting parameters are given relative to
  the 1.0$-$1.5 GeV energy bin.
\label{tbl:SKspecerr}}
\begin{tabular}{c|r|c|c|c}
\hline\hline
  $E_\nu$~(GeV) 
& \multicolumn{1}{|c|}{\ \ \ $\Phi_{\rm{ND}}$\ \ \ }
& $\Delta(\Phi_{\rm{ND}})$
& $\Delta$(F/N) 
& $\Delta(\epsilon_{\rm{SK}})$ \\
\hline
  0$-$0.5    &         1.31\ \   &  49    &   2.6  &  8.7  \\
 0.5$-$0.75  &         1.02\ \   &  12    &   4.3  &  4.3  \\
0.75$-$1.0   &         1.01\ \   &   9.1  &   4.3  &  4.3  \\
 1.0$-$1.5   &  $\equiv1.00$\ \  &   ---  &   6.5  &  8.9  \\
 1.5$-$2.0   &         0.95\ \   &   7.1  &  10    &  10   \\
 2.0$-$2.5   &         0.96\ \   &   8.4  &  11    &  9.8  \\
 2.5$-$3.0   &         1.18\ \   &  19    &  12    &  9.9  \\
 3.0$-$      &         1.07\ \   &  20    &  12    &  9.9  \\
\hline\hline
\end{tabular}
\end{table}
The uncertainty due to neutrino interaction models is separately 
studied. In QE scattering, the axial vector mass in the dipole formula 
is set to a central value of 1.1~${\rm GeV}/c^2$ and is
varied by $\pm$10\%.  The axial mass for single pion production is set
to a central value of 1.2~${\rm GeV}/c^2$ and is varied by 
$\pm$20\%~\cite{Bernard:2001rs}.  
This affects both the $q^2$ dependence of the cross-section and the
total cross-section.  
For coherent pion production, two different models are compared: one
is the Rein and Sehgal model~\cite{Rein:1983pf}, and the other is a
model by Marteau \cite{Marteau:2000ne}.  For deep inelastic
scattering, GRV94~\cite{Bluck:1995uf} and the corrected structure
function by Bodek and Yang~\cite{Bodek:2002md} are studied.  For the
oscillation analysis the Marteau model and Bodek and Yang structure
functions are employed.  Varying the choice of models causes the
fitted value of \Rnqe~($=0.93$) to change by $\sim 20\%$.  In order to
account for this, an additional systematic error of $\pm 20\%$ on
\Rnqe is added to the error matrix.  The choice of models and axial
mass does not affect the spectrum measurement itself beyond the size
of the fitted errors.

%%% Local Variables: 
%%% mode: latex
%%% TeX-master: "oscillation"
%%% End: 

%\section{The measurement of the F/N flux ratios}
%\label{sec:far-near}

The F/N ratio from the beam simulation is used to extrapolate the
measurements at the ND to those at SK.
The errors including correlations above 1~GeV, where the PIMON is
sensitive, are estimated based on the PIMON measurements.  The errors
on the ratio for $E_\nu$ below 1~GeV are estimated based on the
uncertainties in the hadron production models used in the K2K beam
MC~\cite{Ahn:2001cq}.
The diagonal elements in the error matrix for the F/N ratio are
summarized in Table~\ref{tbl:SKspecerr}.
%
%
%

%%% Local Variables: 
%%% mode: latex
%%% TeX-master: "oscillation"
%%% End: 

%\section{Oscillation Analysis}

The events in SK are selected using the timing information provided by
the global positioning system.  Events detected in SK must occur
within an expected beam arrival time window of $1.5$~$\rm \mu sec$.
In addition, the detected events must have no activity in outer
detector, and have an energy deposit greater than 30~MeV with a vertex
reconstructed within the 22.5~kiloton fiducial volume
\cite{Ahn:2001cq}.  This sample of events is referred to as the fully
contained~(FC) sample.  The efficiency of this selection is 93\% for 
CC interactions.  
Fifty-six events satisfy the criteria.  With the timing
cut, the expected number of atmospheric neutrino background is
approximately $10^{-3}$ events.

The expected number of FC events at SK without oscillation is
estimated to be $80.1^{+6.2}_{-5.4}$.
The correlations between energy bins
from the spectrum measurement at the ND and the F/N ratio are taken
into account in the estimation of the systematic errors.  The major
contributions to the errors come from the
uncertainties in the F/N ratio ( $^{+4.9\%}_{-5.0\%}$) and the
normalization (5.0\%), dominated by uncertainties of the fiducial
volumes due to vertex reconstruction both at the 1KT and
SK.

A two flavor neutrino oscillation analysis, with \numu disappearance,
is performed by the maximum-likelihood method.  In the analysis, both
the number of FC events and the energy spectrum shape for $1R \mu$
events are used.  The likelihood is defined as ${\cal L} = {\cal
  L}_{norm} \times {\cal L}_{shape}$.  The normalization term ${\cal
  L}_{norm}(N_{obs}, N_{exp})$ is the Poisson probability to observe
$N_{obs}$ events when the expected number of events is $N_{exp}(\Delta
m^2, \sin^22\theta, f)$.  The symbol $f$ represents a set of
parameters constrained by the systematic errors.  These parameters are
described in detail later.  The shape term, $ {\cal
  L}_{shape}=\prod_{i=1}^{N_{1R\mu}} P(E_i; \Delta m^2, \sin^2 2
\theta, f), $ is the product of the probability for each $1R \mu$
event to be observed at $E_\nu^{\rm rec}=E_i$, where $P$ is the normalized
$E_\nu^{\rm rec}$ distribution estimated by MC simulation and $N_{1R\mu}$
is the number of $1R \mu$ events.

In the oscillation analysis, the whole data sample is used for 
${\cal L}_{norm}$, i.e. $N_{obs}=56$.  The spectrum shape in June
1999 was different from that for the rest of the running period because
the target radius and horn current were different.  The estimation of
energy correlations in the spectrum at the ND and in the far/near
ratio has not been completed for this period.  
Thus, data taken in June 1999 are discarded for ${\cal L}_{shape}$.
The discarded data correspond to 6.5\% of total POT.  
The number of $1R\mu$ events observed excluding the data of June 1999 is 29,
and the corresponding number of $1R\mu$ events expected from MC simulation
in the case of no oscillation is 44.

The parameters $f$ consist of the re-weighted neutrino spectrum
measured at the ND ($\Phi_{ND}$), the F/N ratio, the reconstruction
efficiency ($\epsilon_{SK}$) of SK for $1R \mu$ events, the
re-weighting factor for the QE/non-QE ratio \Rnqe, the SK energy
scale and the overall normalization.  The errors on the first 3 items
depend on the energy and have correlations between each energy bin.  
The diagonal parts of their error
matrices are summarized in Table~\ref{tbl:SKspecerr} as described earlier.
The error on the SK energy scale is 3\%~\cite{Fukuda:1998tw}. 
Two different approaches are taken for the treatment of systematic
errors in the likelihood.  The first is to treat the parameters $f$ as
fitting parameters with an additional constraint term in the
likelihood (method 1)~\cite{Fukuda:1998mi}.  The other approach is to
average the ${\cal L}(f)$ sampled over many random trials weighted
according to the probability density distribution of the systematic
parameters $f$~(method 2)~\cite{Swain:1998cd}.

\begin{figure}[htbp!]
  \centering
  \includegraphics[width=8.0cm]{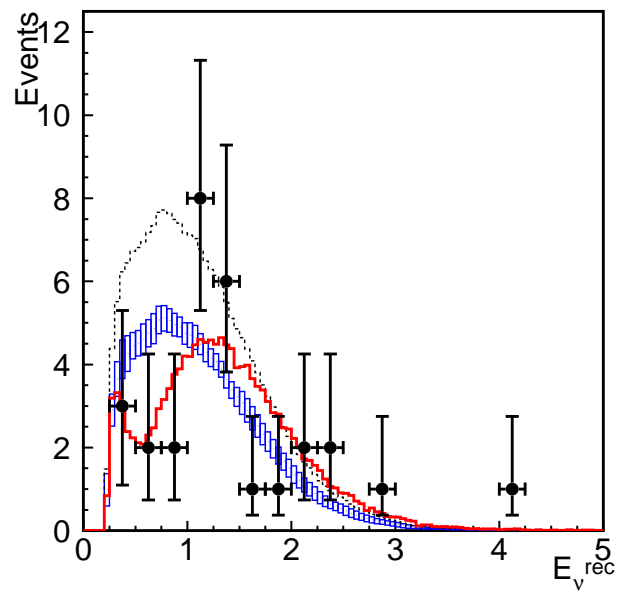} 
  \caption{The reconstructed $E_\nu$ distribution
    for $1R\mu$ sample (from method 1). 
    Points with error bars are data. Box histogram is expected
    spectrum without oscillations, where the height of the box is the
    systematic error. The solid line is the best fit spectrum. These
    histograms are normalized by the number of events observed (29).  In
    addition, the dashed line shows the
    expectation with no oscillations normalized to the expected number of
    events (44).}
  \label{fig:shape}
\end{figure}

%%%%%%%%%%%%%%%%%%%%%%%%%%%%%%%%%%%%%%%%%%%%%
%    BEST FIT
%%%%%%%%%%%%%%%%%%%%%%%%%%%%%%%%%%%%%%%%%%%%%
The likelihood is calculated at each point in the \dms and \sstt space
to search for the point where the likelihood is maximized.  
The best fit point in the physical region of oscillation parameter
space is found to be at (\sstt, \dms)=$(1.0, 2.8\times 10^{-3}~{\rm eV^2})$
in method 1 and at $(1.0, 2.7\times 10^{-3}~{\rm eV^2})$ in method 2.  If
the whole space including the unphysical region is considered the
values are $(1.03, 2.8\times 10^{-3}~{\rm eV^2})$ in method 1 and $(1.05,
2.7\times 10^{-3}~{\rm eV^2})$ in method 2.  The results from two methods
are consistent with each other.  At the best fit point in the physical
region the total number of predicted events is 54.2, which agrees with
the observation of 56 within statistical error.  The observed
$E_\nu^{\rm rec}$ distribution of the $1R \mu$ sample is shown in
Figure~\ref{fig:shape} together with the expected distributions for the
best fit oscillation parameters, and the expectation without oscillations.  
Consistency between the observed and best-fit
$E_\nu^{\rm rec}$ spectrum is checked by using
Kolgomorov-Smirnov(KS) test. A KS probability of 79\% is obtained.
The best fit spectrum shape agrees with the observations.

\begin{figure}[htbp!]
  \centering
   \includegraphics[width=8.0cm]{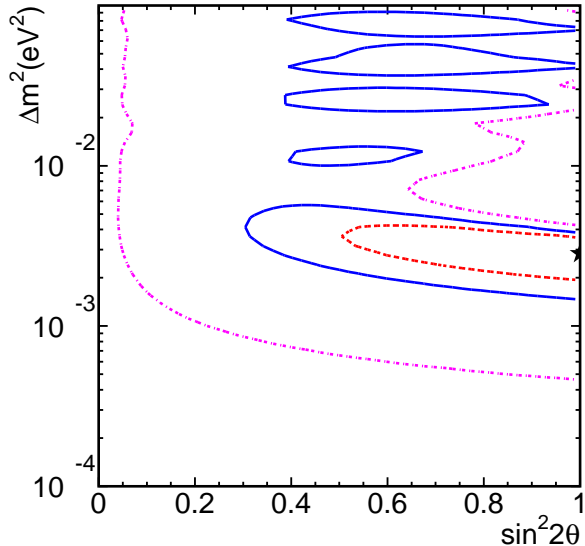} 
  \caption{Allowed regions of oscillation parameters. 
     Dashed, solid and dot-dashed lines are 68.4\%, 90\% and
     99\% C.L. contours, respectively.
    The best fit point is indicated by the star.    
}
  \label{fig:cont}
\end{figure}

%%%%%%%%%%%%%%%%%%%%%%%%%%%%%%%%%%%%%%%%%%%%%
%    NULL OSCILLATION PROBABILITY
%%%%%%%%%%%%%%%%%%%%%%%%%%%%%%%%%%%%%%%%%%%%%
The probability that the observations are due to a statistical
fluctuation instead of neutrino oscillation is estimated by computing
the likelihood ratio of the no-oscillation case to the best fit point.
The no-oscillation probabilities are calculated to be 0.7\% and 0.4\%
for method 1 and 2 respectively. 
When only normalization (shape) information is used, the probabilities
are 1.3\% (16\%) and 0.7\% (14.3\%) for the two methods.  
%%%%%%%%%%%%%%%%%%%%%%%%%%%%%%%%%%%%%%%%%%%%%
%    ALLOWED REGION
%%%%%%%%%%%%%%%%%%%%%%%%%%%%%%%%%%%%%%%%%%%%%
Allowed regions of oscillation parameters are evaluated by calculating
the likelihood ratio of each point to the best fit point, and are
drawn in Figure~\ref{fig:cont}.  Both methods give essentially the same
results.  In order to be conservative, the result from method 1 is
shown in the figure as it gives a slightly larger allowed region at
the 99\% C.L.  The 90\% C.L.  contour crosses the $\sin^22\theta=1$
axis at $1.5$ and $3.9\times 10^{-3}$~eV$^2$ for \dms.
The oscillation parameters preferred by the total flux suppression and
the energy distortions alone also agree well.  Finally, the
uncertainties of neutrino interactions are studied using the same
procedure as the spectrum measurement at the ND.  It is found that the
effects of the interaction model difference on all the results are
negligible due to the cancellation caused by using the same models in
both the ND and SK.

%%% Local Variables: 
%%% mode: latex
%%% TeX-master: "oscillation"
%%% End: 

%\section{conclusions}
%\label{sec:conclusions}

In conclusion,
%the K2K experiment conducted a search for neutrino
%oscillations by using accelerator produced neutrinos to investigate
%the oscillation phenomena discovered in atmospheric neutrinos.  The
%neutrino flux at SK is predicted based on the measurements at the near
%detectors and the extrapolation from the near detectors to SK by beam MC
%simulations.  These simulations are validated by the measurements of
%the pion monitor.  
both the number of observed neutrino events and the observed energy
spectrum at SK are consistent with neutrino oscillation.
The probability that the measurements at SK are explained by statistical
fluctuation is less than 1\%.  The measured oscillation parameters are
consistent with the ones suggested by atmospheric neutrinos.
%, and the measurement of \dms at \sstt = 1.0 is
%one of the most precise measurements.  
At the time of this letter the K2K experiment has collected
approximately one-half of its planned $10^{20}$ protons on target.

% to provide
%sufficient statistics for the further study of neutrino oscillation.

%%% Local Variables: 
%%% mode: latex
%%% TeX-master: "oscillation"
%%% End: 

\begin{acknowledgments}
  We thank the KEK and ICRR Directorates for their strong support and
  encouragement.  K2K is made possible by the inventiveness and the
  diligent efforts of the KEK-PS machine and beam channel groups.  
  We gratefully
  acknowledge the cooperation of the Kamioka Mining and Smelting
  Company.  This work has been supported by the Ministry of Education,
  Culture, Sports, Science and Technology, Government of Japan and its
  grants for Scientific Research, the Japan Society for Promotion of
  Science, the U.S. Department of Energy, the Korea Research
  Foundation, and the Korea Science and Engineering Foundation.
\end{acknowledgments}

\bibliography{references}
%\footnote{Present address can be found on http://neutrino.kek.jp/presentaddress0211.giff}
\end{document}